# Room Temperature Nanoparticulate Interfacial Layers for Perovskite Solar Cells via solvothermal synthesis

Achilleas Savva[a], Ioannis T. Papadas[a], Dimitris Tsikritzis[a], Gerasimos S. Armatas[b], Stella Kennou[c] and Stelios A. Choulis[a,*]

[a]Molecular Electronics and Photonics Research Unit, Department of Mechanical Engineering and Materials Science and Engineering, Cyprus University of Technology, Limassol, 3603, Cyprus.

[b]Department of Materials Science and Technology, University of Crete, Heraklion 71003, Greece.

[c]Department of Chemical Engineering, University of Patras, Patra, 26504 Greece

* stelios.choulis@cut.ac.cy

We present a solvothermal synthetic route to produce monodispersed CuO nanoparticles (NPs) in the range of 5-10 nm that can be used as hole selective interfacial layer between indium tin oxide (ITO) and perovskite active layer for p-i-n perovskite solar cells by a spin casting the dispersions at room temperature. The bottom electrode interface modification provided by spherical CuO-NPs at room temperature promotes the formation of high quality perovskite photoactive layers with large crystal size and strong optical absorption. Furthermore, it is shown that the nanoparticulate nature of the CuO hole transporting interfacial layer can be used to improve light manipulation within perovskite solar cell device structure. The corresponding p-i-n $CH_3NH_3PbI_3$-based solar cells show high Voc values of 1.09 V, which is significantly higher compared to the Voc values obtained with conventional PEDOT:PSS hole selective contact based perovskite solar cells.

## Introduction

Organic-inorganic lead halide perovskites have been identified as key materials for many of optoelectronic applications.[1-2] Perovskite-based solar cells have demonstrated power conversion efficiencies (PCEs) above 20%[3-4] within only few years of development.[5] Solid-state perovskite solar cells have been triggered by the pioneered work of H.-S. Kim *et al*.[6] Since then, perovskite solar cell device engineering has made substantial improvement towards facile, innovative concepts. Planar heterojunction (PHJ) perovskite-based solar cells being processed by cost-effective solution routes highlight the great merit for future commercialization.[7]

Inverted p-i-n perovskite solar cells have recently attracted increased interest[8] due to low-cost and easy processing of the charge selective contacts that can be used in this device structure. Importantly within the p-i-n perovskite the utilization of high cost hole transporting layers like Spiro-MeOTAD as well as high temperature processed n-type materials like $TiO_2$ at 500 °C can be avoided. In addition, the fullerene based interfacial layer processed on top of the perovskite as a n-type contact has been found to reduce the current-voltage (J-V) curve hysteresis which is generally observed in n-i-p structures when applying different scan directions.[9]

In principle, the corresponding bandgap ($E_g$) of the photoactive material determines the upper limit of the achievable open-circuit voltage (Voc) in planar heterojunction (PHJ) perovskite solar cells. According to the Shockley-Queisser theory,[10] there are intrinsically thermodynamic constraints for materials that eventually result in a minimum voltage loss of about 250-300 mV. Moreover, charge recombination within the photoactive material and energetic charge barriers in the stratified devices may also contribute to an additional voltage loss, further reducing the resultant Voc of the device.[11] Importantly, perovskite solar cells show the potential of achieving low voltage loss (< 0.5 V).[12] Therefore, it is critical to pursue material systems with minimal voltage loss for developing high-performance solar cell devices.

Poly(3,4-ethylenedioxythiophene) polystyrene sulfonate (PEDOT:PSS) is a widely-used p-type material for optoelectronic applications[13] and the most commonly used hole selective contact in p-i-n perovskite solar cells. However, despite its excellent properties, most of the PEDOT:PSS-based $CH_3NH_3PbI_3$ solar cells reported to date suffered an inferior open-circuit voltage ($V_{OC}$ ~ 0.88-0.97 V) compared with that obtained for conventional $CH_3NH_3PbI_3$-based solar cells (1.05-1.12 V).[14] It is generally believed that the lower Voc originates from the mismatched energy level alignment between the work function of PEDOT:PSS (4.9-5.2 eV)[15] and the valence band maximum (5.4 eV) of $CH_3NH_3PbI_3$. Furthermore, it is observed that the perovskite photoactive layers, which are formed on top of the PEDOT:PSS, exhibit relatively small grain sizes that significantly deteriorate the overall device performance.

In general, it is proved that the under-layers have a significant influence on the grain size of the perovskite photoactive layers which are formed on top.[16] Polymeric p-type materials, like poly[bis(4-phenyl) (2,4,6-trimethylphenyl)amine] (PTAA), have been reported as efficient hole selective contacts, leading to larger perovskite grain sizes compared with those formed on top of PEDOT:PSS. In addition, the higher work function of these materials results in enhanced bottom electrode selectivity and, thus, the formation of perovskite solar cells with higher Voc. Nevertheless, the high cost of these materials due to their complicated synthetic procedure and high-purity requirement, immensely hampering the future commercialization of perovskite solar cells.

For further cost-effective application, low-cost, simply prepared and stable inorganic hole conductors have been identified as promising alternatives to the costly organic hole conductors and low performing PEDOT:PSS. Solution processed NiO,[17] CuNiOx,[18] and CuSCN[19] materials have been reported as efficient hole selective contacts, which result in high Voc values (> 1 V) and increased p-i-n device performance compared with PEDOT:PSS-based perovskite solar cells.

Typically, to form functional solution processed metal oxides as hole-transport layers (HTLs) for p-i-n perovskite solar cells, modified sol-gel methods are used. These methods usually require coating of a precursor metal salt solution and then a post-annealing treatment in order to achieve high quality, electrically active metal oxide thin films.[20] The high temperature processing steps involved in this process, however, are not compatible with the flexible perovskite solar technology targets.[21] A second approach for fabricating metal oxide based HTLs for p-i-n perovskite solar cells is the synthesis of high crystallinity and high purity nanoparticle (NP) dispersions.[22-23] The dispersions are processed using plain solution processing followed by post deposition treatments to remove excess organic residuals and to promote film functionality.[24] It has been reported that this approach could provide more flexibility in processing by avoiding prolonged post deposition treatments.[25]

For efficient planar perovskite solar cells application, the compactness of the inorganic nanoparticulate under layers is of high importance.[26] To achieve this, the synthesized NPs are crucial to have small size and narrow distribution of the particle sizes.[27] A typical problem of the existing synthetic routes is that the as-prepared NPs are severely agglomerated, thus making it difficult to control the size and shape of the particles. In addition, in most of the cases these methods require the usage of toxic and expensive reagents or proceed at high temperature and through complex synthetic routes.

CuO is a low-cost material which has been used in various applications, such as catalysis,[28] water splitting photo catalysis[29-30] and gas sensing.[31] CuO is a promising HTL for perovskite solar cells.[32] Recent studies demonstrate CuOx-based HTLs starting from metal salts precursor[33-34] and Cu(acac)$_2$ compound in o-DCB solution[35]. All the reported CuOx-HTLs require thermal-annealing steps combined in some cases with additional solvent washing steps to ensure HTL functionality.

In this study, we demonstrate a new strategy for fabricating nanoparticulate interfacial layers for p-i-n perovskite solar cells based on copper(II) oxide NP dispersions. Importantly for printed electronic applications the proposed synthetic root can be used to develop functional CuO HTL at room temperature eliminating the requirements of annealing step during the fabrication process. Furthermore, we show that the proposed nanoparticulate based interface modification concept can be used to improve light manipulation providing in parallel a series of other benefits in p-i-n perovskite solar cell device operation. The cupric oxide (CuO) NPs are synthesized from CuCl thermal decomposition in DMSO at low temperature (80-120 °C) and within short reaction time (30 minutes). As proved by XRD, TEM and EDS measurements, the CuO NPs are of high purity with monoclinic crystal structure, spherical in shape and with tunable diameter (from 5 to 10 nm) depending on the reaction conditions. Stable dispersions of the CuO NPs in DMSO were used to fabricate interfacial layers of CuO without any post deposition annealing or solvent washing step in the range of ~15 nm. The room temperature solution processed CuO interfacial layers exhibit high work function, transparency and good surface properties to facilitate high quality perovskite photoactive layers on top of them with significantly increased grain size compared with PEDOT:PSS. It is proved that the use of CuO NPs as interfacial layers between ITO and $CH_3NH_3PbI_3$ leads in good photovoltaic operation with maximum PCE of 15.3%, high Voc values of 1.09 V and over 50% increase in PCE compared with PEDOT:PSS-based p-i-n perovskite-based solar cells. The devices under study are analyzed using electro-optical characterization suggesting that the nanoparticulate nature of the CuO interfacial layer is leading in a light management benefits for the photovoltaic device performance of the p-i-n perovskite solar cells.

**Experimental**

*Materials:* Pre-patterned glass-ITO substrates (sheet resistance 4Ω/sq) were purchased from Psiotec Ltd, $PbI_2$ from Alfa Aesar, MAI from Dyenamo Ltd, Aluminum-doped zinc oxide (AZO) ink from Nanograde (product no N-20X), PC[70]BM from Solenne BV and PEDOT:PSS from H.C. Stark (Clevios P VP Al 4083). All the other chemicals used in this study were purchased from sigma Aldrich.

*Synthesis of CuO NPs:* 2 mole of copper chloride were added into 10 mL of dimethyl sulfoxide (DMSO). The solution was heated at various temperatures and desired reaction time while constantly stirring it at a high rate, thus forming a homogeneous solution. The formation of CuO NPs was indicative by the color change of the precursor solution to black. To clean the CuO NPs from impurities and to improve the quality of the dispersions, we used the followed procedure: the CuO NPs were isolated by centrifugation, washed with DMSO/toluene (2:1 v/v) twice, and re-dispersed in DMSO. Three series of CuO NPs were synthesized under the scope of this study by changing only the decomposition temperature, i.e. CuO-20 mg/mL-120 °C-30 minutes, CuO-20 mg/ml-100 °C-30 minutes and CuO-20 mg/ml-80 °C-30 minutes. We also found that the size of NPs can be increased by using longer reaction times or more concentrated precursor solutions (data not shown).

*Device Fabrication:* The p-i-n solar cells under comparison was ITO/PEDOT:PSS or CuO-NPs/$CH_3NH_3PbI_3$/PC[70]BM/AZO/Al. ITO substrates were sonicated in acetone and subsequently in isopropanol for 10 minutes and heated at 120 °C on a hot plate 10 minutes before use. To form a 40 nm hole transporting layer, the PEDOT:PSS ink was dynamically spin coated at 6000 rpm for 30s in air on the preheated ITO substrates and then transferred to a nitrogen filled glove box for a 25 minutes annealing at 120 °C. To form a ~15 nm CuO nanoparticulate thin film the as prepared dispersions were diluted to 0.5 mg/ml in DMSO and dynamically spin coated at 6k rpm followed by a methanol washing step at 3k-30sec-30ul. The perovskite solution was prepared 30 minutes prior spin coating by mixing $PbI_2$:Methylamonium iodide (1:3) at 46% wt in DMF with the addition of 1.5% mole of methylamonium bromide (to $PbI_2$). The precursor was filtered with 0.1 μm PTFE filters. The perovskite precursor solution was deposited on the HTLs by static spin coating at 4k rpm for 60 seconds and annealing for 5 minutes at 80 °C, resulting in a film with a thickness of ~200 nm. The PC[70]BM solution, 20 mg/ml in chlorobenzene, was dynamically spin coated on the perovskite layer at 1k rpm for 30 s. AZO ink was dynamically spin coated on top of PC[70]BM at 3k rpm for 30 s to fabricate ~30 nm thin films. Finally, 100 nm Al layers were thermally evaporated through a shadow mask to finalize the devices. Encapsulation was applied directly after evaporation in the glove box using a glass coverslip and an Ossila E131 encapsulation epoxy resin activated by 365 nm UV-irradiation. The active area of the devices was 0.09 mm$^2$.

*Characterization:* The thicknesses of the active layers were measured with a Veeco Dektak 150 profilometer. The current density-voltage (J/V) characteristics were characterized with a Botest LIV Functionality Test System. Both forward (short circuit -> open circuit) and reverse (open circuit -> short circuit) scans were measured with 10 mV voltage steps and 40 ms of delay time. For illumination, a calibrated Newport Solar simulator equipped with a Xe lamp was used, providing an AM1.5G spectrum at 100 mW/cm$^2$ as measured by a certified oriel 91150V calibration cell. A custom-made shadow mask was attached to each device prior to measurements to accurately define the corresponding device area. Optical absorption measurements were performed using a Shimadzu UV-2700 UV-Vis spectrophotometer. Diffuse reflectance was recorded at room temperature, using powder $BaSO_4$ as a 100% reflectance standard. Reflectance data were converted to absorption ($\alpha/S$) according to the Kubelka-Munk equation: $\alpha/S = (1-R)^2/(2R)$, where R is the reflectance and $\alpha$, $S$ are the absorption and scattering coefficients, respectively. Steady-state PL experiments were performed in a Fluorolog-3 Horiba Jobin Yvon spectrometer based on an iHR320 monochromator equipped with a visible photomultiplier tube (Horiba TBX-04 module). The PL was non-resonantly excited at 400 nm by the line of 5 mW Oxxius laser diode. Contact angle measurements were performed using a KRUSS DSA100E drop shape analysis system. The value of the water contact angle was calculated by the EasyDrop software, using the sessile drop method. For the surface energy measurements, the contact angles of 3 known surface tension liquids were measured on top of the surfaces under study. The polar and dispersive parts of the surface tension of the liquids were calculated via pendant drop method. Atomic force microscopy (AFM) images were obtained using a Nanosurf easy scan 2 controller under the tapping mode. X-ray diffraction (XRD) patterns were collected on a PANanalytical X´pert Pro MPD powder diffractometer (40 kV, 45 mA) using Cu Kα radiation (λ=1.5418 Å). Transmission electron microscope (TEM) images and electron diffraction patterns were recorded on a JEOL JEM-2100 microscope with an acceleration voltage of 200 kV. The samples were first gently ground, suspended in ethanol and then picked up on a carbon-coated Cu grid. Quantitative microprobe analysis was performed on a JEOL JSM-6390LV scanning electron microscope (SEM) equipped with an Oxford INCA PentaFET-x3 energy dispersive X-ray spectroscopy (EDS) detector. Data acquisition was performed with an accelerating voltage of 20 kV and 60 s accumulation time. X-ray photoelectron spectra (XPS) and Ultraviolet Photoelectron Spectra (UPS) were recorded by Leybold EA-11 electron analyzer operating in constant energy mode at pass energy of 100 eV and at a constant retard ratio of 4 eV for XPS and UPS respectively. The spectrometer energy scale was calibrated by the Au4f$_{7/2}$ core level binding energy, BE, (84.0 ± 0.1 eV) and the energy scale of the UPS measurements was referenced from the Fermi level position of Au at binding energy of 0 eV. All binding energies were referred to the C 1s peak at 284.8 eV of surface adventitious carbon, respectively. The X-ray source for all measurements was an un-monochromatized Al Kα line at 1486.6 eV (12 keV with 20 mA anode current). For the UPS measurements, the He I (21.22 eV) excitation line was used. A negative bias of 12.22 V was applied to the samples during UPS measurements in order to separate secondary electrons originating from sample and spectrometer. The analyzer resolution was determined from the width of the Au Fermi edge to be 0.16 eV. The sample work function was determined by subtracting from the HeI excitation energy (21.22 eV) the high binding energy cut-off. The position of the high-energy cut-off position was determined by the intersection of a linear fit of the high binding portion of the spectrum with the background. Similarly, the valance band maximum is determined in respect to the Fermi level, from the linear extrapolation of the valence band edge to the background.

**Results and Discussion**

The CuO NPs were synthesized using a simple synthetic procedure at low temperature as described in details within the experimental section. The thermal decomposition of CuCl in DMSO was proven as an uncomplicated way to tune the particle size by simple changing the reaction temperature from 80 to 120 ºC. In this procedure, DMSO acts as ancillary ligand to form complexes with metal halide precursor in a way that molecules of solvent replace $Cl^-$ in CuCl through a bridge-splitting reaction, producing monodisperse NPs of CuO during the thermal decomposition synthesis. Fig. 1 shows XRD and TEM data which accurately illustrate the crystal phase, shape and size of the as-obtained CuO NPs. The initial concentration of the CuCl solution was 20 mg/ml and the reaction time was 30 min in all experiments.

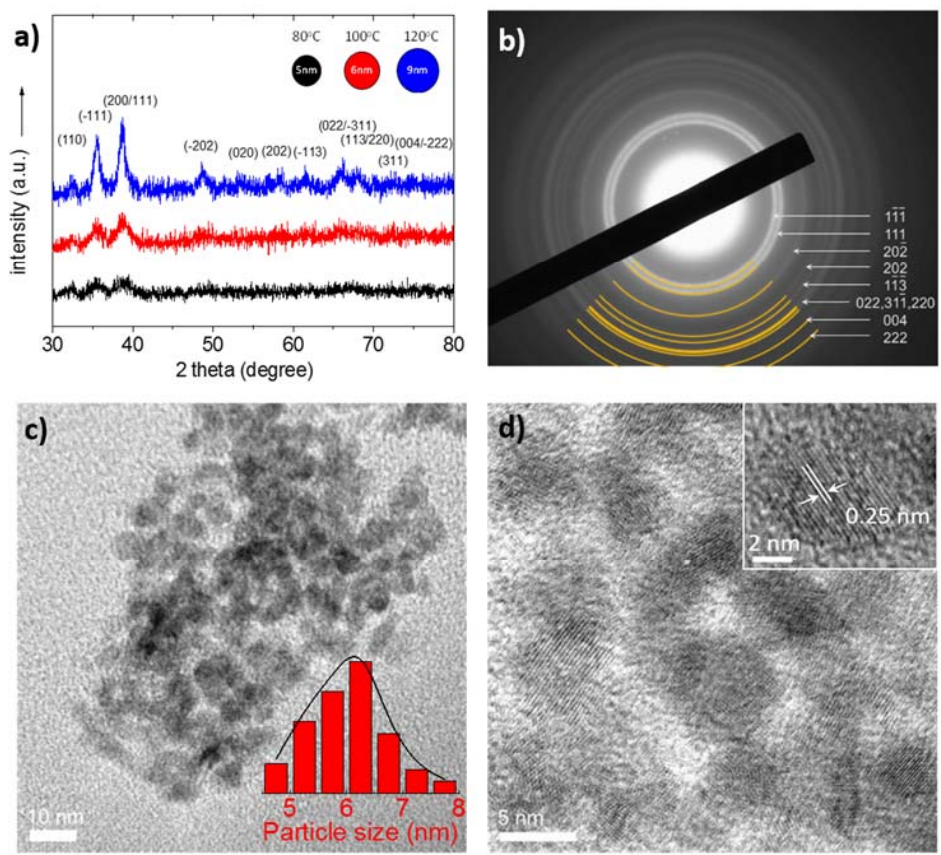

**Fig. 1 a)** XRD patterns of CuO NPs obtained at 80 °C (black), 100 °C (red), 120 °C (blue line) decomposition temperatures. **b)** SAED pattern and **c, d)** representative TEM images for CuO NPs obtained at 100 °C decomposition temperature. The insert in panel c displays the histogram of the size distribution of CuO NPs, showing an average particle diameter of 5.9 ± 0.7 nm. The insert in panel d displays high-resolution TEM (HRTEM) image of an individual CuO NP annealed at 100 °C.

Fig. 1a shows the XRD patterns for the CuO NPs prepared at three different reaction temperatures, i.e. 80, 100 and 120 °C; all the XRD peaks can be assigned to the monoclinic structure of CuO with lattice constants of $a$=0.4688 nm, $b$=0.3423 nm, $c$=0.5129 nm and

*β*=99.506º (JCPDS Card No. 48-1548). No peaks of impurities were observed in the XRD patterns, indicating the high phase purity of the CuO. The latter is also verified by EDS analysis in which very low impurities of Cl have been detected (less than 2% by weight, supporting Fig. S1). The crystal structure of the CuO NPs was further studied by selected-area electron diffraction (SAED). The SAED pattern in Fig. 1b, taken from a small area of the NP aggregates (as formed by TEM preparation of sample), corroborates the crystallinity observed in XRD patterns. It shows a series of Debye-Scherrer diffraction rings, which can be well indexed to monoclinic CuO (space group: C2/c, tenorite). The average size of the CuO NPs was estimated from the (-111) diffraction peak by using the Scherer formula, and it was found to be ~5.7 nm for 80 ºC, ~6.3 nm for 100 ºC and ~10 nm for 120 ºC reaction temperature. Fig. 1c and d show typical TEM images for the CuO NPs prepared at 100 ºC. These images clearly show spherical NPs with an average size of 5.9 ±0.7nm (inset of Fig. 1c), in consistent with XRD results. Note that the average diameter of the CuO NPs was estimated by counting more than 100 individual NPs in several TEM images (Fig. S2). The high resolution TEM (HRTEM) image, in Fig. 1d (inset), shows lattice fringes of a CuO NP, suggesting a well-defined crystal structure. The lattice fringes with d-spacing of 2.5 Å are assigned to the (002) planes of the CuO phase with monoclinic structure. Finally, to investigate the electronic structure of the synthesized CuO NPs, diffuse reflectance UV-Vis-near-IR spectroscopy was used. Fig. S3 represents the Tauc plots for direct band gap transitions, from which the optical energy gap was estimated to be 1.45, 1.43 and 1.38 eV for CuO NPs prepared at 80, 100 and 120 ºC, respectively.

To fabricate compact CuO NPs thin films stable dispersions in appropriate solvents are of high importance. All the CuO NPs produced at different temperatures were well dispersed in DMSO without the use of any dispersive agents (Fig. S4). Sulfoxides, such as DMSO, are proposed to serve as a ligand for binding to the CuO NP surface and thus play a crucial role in stabilizing CuO NPs and promoting the monodispersity of nanocrystals. Characterization of the coordination properties of DMSO in these reactions, however, is complicated and insights thus far have been limited to computational studies. The as-prepared CuO NPs with initial concentration of 20 mg/ml were further diluted down to 0.5 mg/ml for the fabrication of homogeneous layers with 15 nm thick. As shown in AFM image in Fig. S5, compact thin films of CuO NPs can be fabricated by a simple spin coating of the dispersions without any post deposition treatment. However, it is observed that the NPs size has a great influence on the resulting thin films properties. In particular, compact ultrathin films necessary for good photovoltaic operation are not able to be produced by using CuO NPs synthesized at 120 ºC (average particle size ~ 9nm). In addition, despite that compact thin films in the range of ~15 nm can be obtained from CuO NPs synthesized at 80 ºC (average particle size ~ 5nm), we observe significantly lower PCEs for the corresponding devices (Fig. S6). The CuO NPs prepared at 100 ºC (average particle size ~ 6 nm) were found to lead to higher PCEs compared with the other CuO samples and are extensively analyzed in the rest of the letter.

As previously mentioned the electronic and surface properties of the bottom electrode are crucial for device operation and the formation of perovskite photoactive layers. Fig. 2 shows the, XPS/UPS studies, optical absorption spectra and surface wetting envelopes of ITO/CuO NPs (prepared at 100 ºC) films.

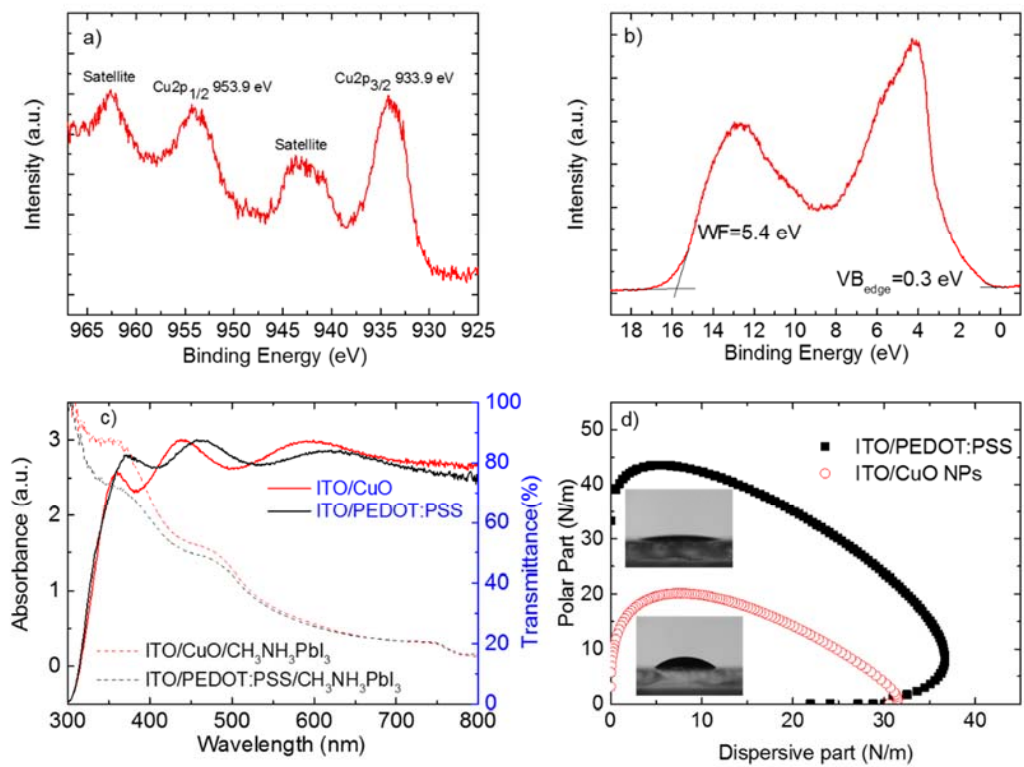

**Fig. 2 a)** The XPS spectrum of the Cu 2p peak, **b)** UPS spectrum of ITO/CuO substrates, **c)** Absorbance spectra of ITO/PEDOT:PSS/CH$_3$NH$_3$PbI$_3$ (black dashed line) and ITO/CuO NPs/CH$_3$NH$_3$PbI$_3$ (red dashed line) electrodes. Transmittance spectra (right axis) of ITO/PEDOT:PSS (black solid line) and ITO/CuO NPs (red solid line), **b)** Surface energy wetting envelopes of ITO/PEDOT:PSS (black filled squares) and ITO/CuO NPs (red open circles). The insets represent digital photographs of a water drop on top of the corresponding layers.

Fig. 2a shows the XPS Cu 2p peak. The Cu $2p_{3/2}$ peak is located at 933.9 eV and is accompanied by a broad satellite on the high binding energy side at about 9-10 eV. Both the binding energy of Cu 2p3/2 and the fine structure of the satellite are characteristic of CuO and are in agreement with literature.[36-37] This is an experimental evidence that high quality CuO layers are formed on top of ITO without any annealing treatment.

Fig. 2b shows the UPS spectrum of CuO/ITO interface. The work function of the CuO was determined at 5.4 eV from the high binding energy cut-off as is descripted in the experimental section. The valance band region near the fermi level shows broad features between 2-7 eV originating from overlapping Cu 3d and $O_2$ p levels.[38] The metallic nature of CuO is confirmed by the occupied states very close to the Fermi level and the valence band maximum was determined at 0.3 eV below the Fermi level. CuO has d-bands partially occupied with a high number of electrons and the valence band maximum is reported at 0.1 eV below the Fermi level, but when the oxide was prepared in ultra-high vacuum conditions.[39] Here, the thin layer of the adventitious carbon present on top of the CuO, conceal the states near the Fermi level. The ionization energy of the CuO was found at 5.7 eV, which is calculated by adding the work function and valence band maximum values.

Fig. 2c shows that the transparency of both ITO/PEDOT:PSS/$CH_3NH_3PbI_3$ and ITO/CuO NPs/$CH_3NH_3PbI_3$ electrodes is above 80% in all the visible spectrum (400-800 nm). This guarantees that many photons reach the $CH_3NH_3PbI_3$ photoactive layer. Importantly, the absorbance spectra of identically fabricated $CH_3NH_3PbI_3$ photoactive layers on top of the two bottom electrodes are significantly different. The absorbance of $CH_3NH_3PbI_3$ on top of CuO NPs is stronger compared with that of $CH_3NH_3PbI_3$ fabricated on top of PEDOT:PSS, in all range of wavelengths (300-500 nm) studied. The high transparency of the ITO/CuO NPs bottom electrode as well as the improved absorption characteristics of the perovskite photoactive layer fabricated on top it, as we shall show bellow, is of great importance for the performance of the corresponding photovoltaic devices.

Another important requirement for the bottom electrode of perovskite solar cells is its surface wetting properties. Fig. 2b shows the surface free energy of the two bottom electrodes under study. Surface energy wetting envelopes are closed curves that are obtained when the polar fraction of a solid is plotted against the disperse part.[40] With the help of the wetting envelope and knowledge of the polar and disperse parts of the surface energy (SFE) of a solid it is possible to predict whether a liquid whose surface tension components lie within this enclosed area will wet the corresponding solid. As shown in Fig. 2b, the calculated surface tension of the $CH_3NH_3PbI_3$ precursor solution lies well inside the surface energy wetting envelope of the ITO/CuO bottom electrode and thus a good wetting is guaranteed. In addition, it is observed that the ITO/CuO NPs film exhibits a smaller size of surface wetting envelope than the ITO/PEDOT:PSS thin film, which implies more hydrophobic character. The increased hydrophobicity of the under layer is very possible to lead in perovskite photoactive layer with increase grain size as explained in details elsewhere.[16]

To verify the surface topography of the $CH_3NH_3PbI_3$ photoactive layers fabricated identically on top of ITO/PEDOT:PSS and ITO/CuO NPs thin films, the as-obtained devices were analyzed by AFM measurements which are demonstrates in Fig. 3 together with the steady state photoluminescence (PL) measurements.

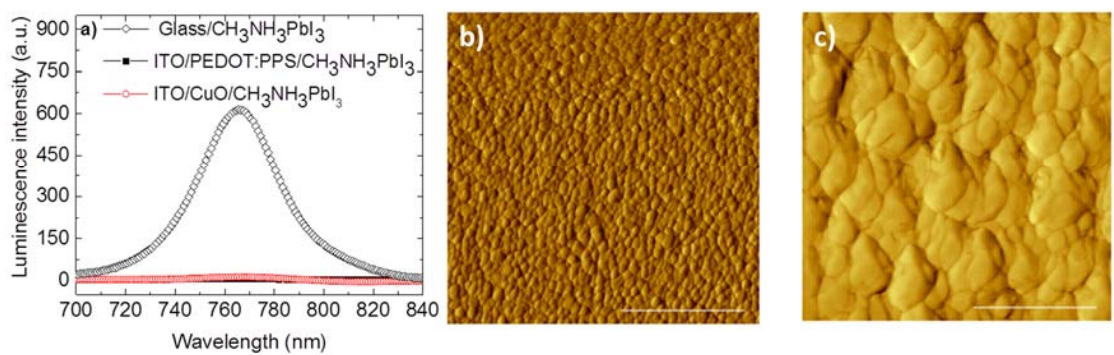

**Fig. 3 a)** Photoluminescence spectra of glass/CH$_3$NH$_3$PbI$_3$ (black open diamonds) ITO/PEDOT:PSS/CH$_3$NH$_3$PbI$_3$ (black filled squares), ITO/CuO/CH$_3$NH$_3$PbI$_3$ (red open circles). AFM measurements of **b)** ITO/PEDOT:PSS/CH$_3$NH$_3$PbI$_3$ and **c)** ITO/CuO/CH$_3$NH$_3$PbI$_3$. The scale bar is 2 μm.

Fig. 3a shows the steady state PL spectra of CH$_3$NH$_3$PbI$_3$ fabricated on top of glass and on top of ITO/CuO or ITO/PEDOT:PSS. The PL spectrum of CH$_3$NH$_3$PbI$_3$ fabricated on top of glass shows a strong pick at 764 nm. This strong signal is completely quenched when the CH$_3$NH$_3$PbI$_3$ layer is fabricated on ITO/CuO as well as ITO/PEDOT:PSS with only few counts measured representing nearly 100% PL quenching. The 100% PL quenching of the main CH$_3$NH$_3$PbI$_3$ PL peak at 764 nm clearly demonstrates excellent ITO/CuO bottom electrode operation.

Fig. 3b demonstrates the surface topography of CH$_3$NH$_3$PbI$_3$ fabricated on top of ITO/PEDOT:PSS electrodes. It is observed that the CH$_3$NH$_3$PbI$_3$ film is quite compact and relatively smooth (Ra = 12.2 nm). However, the grain size of this film is very small and falls within the range of ~100 nm. Importantly, an impressive increase of CH$_3$NH$_3$PbI$_3$ grains size is observed (Fig. 3c) when fabricated on top of ITO/CuO NPs. In this case, an average grain size of ~500 nm is observed and the film maintains its compactness with a surface roughness Ra ~ 18 nm. The impressive increase of the perovskite grain size when fabricated on top of CuO NPs could be partially attributed to the increased hydrophobicity of CuO NPs compared with PEDOT:PSS (Fig. 2b). Indeed, this observation is supported by the results obtained from optical absorbance measurements (Fig. 2a). The increased absorbance of the perovskite photoactive layer on ITO/CuO NPs is a result of the increased perovskite grain size, which lead to an increased surface coverage and stronger absorption form the $CH_3NH_3PbI_3$ crystals.

So far, we have analyzed the nature of the synthesized NPs and showed that the corresponding thin films fulfill a lot of the requirements for serving as an efficient bottom electrode in p-i-n perovskite solar cells. Therefore, p-i-n devices comprised of ITO/CuO NPs/$CH_3NH_3PbI_3$/PC[70]BM/AZO/Al (CuO-device) were fabricated and compared with ITO/PEDOT: PSS/$CH_3NH_3PbI_3$/PC[70]BM/AZO/Al (PEDOT: PSS-device). Fig. 4 shows the photovoltaic operation of the best p-i-n devices under study using current-voltage plots (J-V) and external quantum efficiency (EQE). The results are presented in Fig. 4 and were verified in several other identically executed experimental runs.

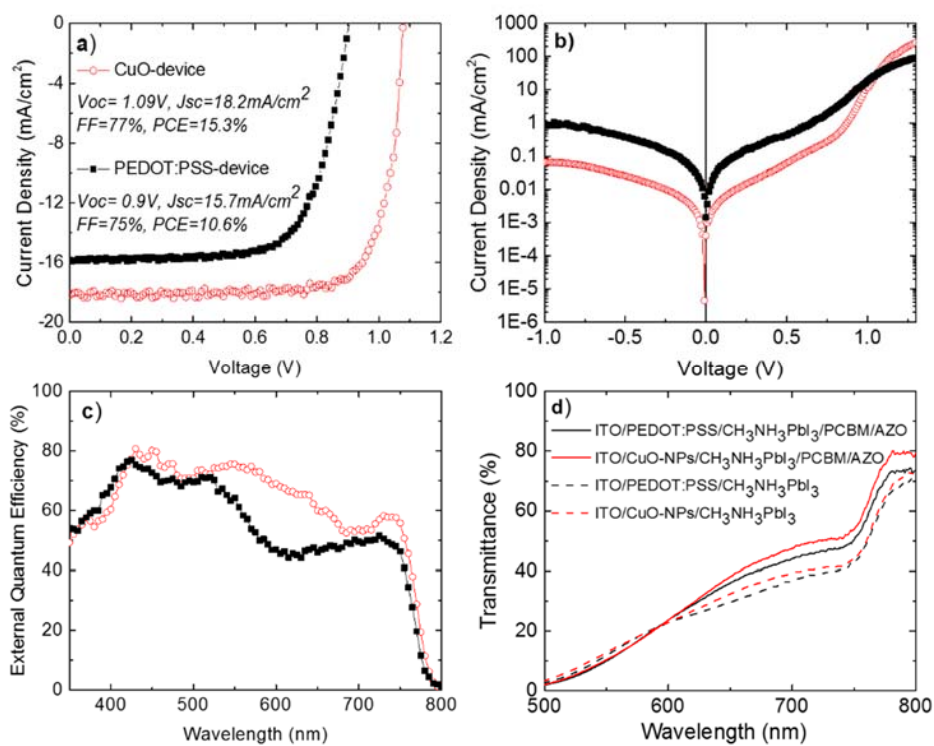

**Fig. 4** Current density–Voltage (J-V) characteristics **a)** under illumination, **b)** under dark conditions, **c)** EQE spectra for the two p-i-n devices under this study ITO/PEDOT: PSS/CH$_3$NH$_3$PbI$_3$/PC[70]BM/AZO/Al (black filled squares) and ITO/CuO NPs/CH$_3$NH$_3$PbI$_3$/PC[70]BM/AZO/Al (open red circles) and **d) t**ransmittance measurements of different layer stacks comprising the p-i-n solar cells under study.

Fig. 4a shows the J-V plots under illumination for the best p-i-n devices under study. Both CuO-based and PEDOT:PSS-based devices exhibited good diode behavior with high fill factor (FF), open-circuit voltage (Voc) and short circuit current density (Jsc) values. It is worthy to mention that both p-i-n devices show negligible hysteresis between the two-different scan J-V directions (Fig. S7). Importantly, the photovoltaic performance of CuO-based perovskite solar cell presented is stable under for more than 30 minutes under continues 1 sun illumination (Fig. S8), indicating no photo-chemical interaction between CuO HTL and $CH_3NH_3PbI_3$. The PEDOT:PSS-based device exhibited Voc = 0.9 V, Jsc = 15.7 mA/cm$^2$, FF = 75% and PCE = 10.6%. Importantly, CuO NPs-based inverted device exhibited Voc = 1.09 V, which is 0.19 V higher compared to that of PEDOT:PSS-based device (Jsc = 18.2 mA/cm$^2$, FF = 77% and PCE = 15.3%). In addition to best device performance presented in Fig. 4, CuO-based p-i-n perovskite solar cells exhibit significant average device performance improvement compared with PEDOT:PSS-based devices and good performance reproducibility as shown in statistical analysis in Fig. S9. As predicted before, we ascribe the enhanced Voc to the higher work function of CuO (5.4 eV, Fig. 2b), than that of PEDOT:PSS (5 eV).[15] The latter leads in increased built-in-voltage (Vbi) as predicted by the p-i-n model and thus increased Voc. The higher work function is most possibly responsible for the increased Jsc and FF for the CuO-device by providing better energetic steps and thus reducing energetic barriers for unimpeded hole collection for the bottom electrode. In addition, the increased $CH_3NH_3PbI_3$ grain size when fabricated on top of CuO NPs compared to those fabricated on top of PEDOT:PSS (Figs. 3b and c) could also have a merit on the increased photovoltaic parameters for the CuO-based device. The increased grain size of $CH_3NH_3PbI_3$ photoactive layer would lead in reduced grain boundaries area and thus reduced non-geminate recombination probability, which is a parasitic process dominated at the grain boundaries and limiting the Voc and Jsc values of perovskite solar cells.[41]

The above described mechanisms are further verified by Fig. 4b. CuO-NPs based device exhibits a very good diode operation. Importantly, the leakage current (at voltage equal to -1 V) for the CuO NPs based device is 1 order of magnitude lower compared with PEDOT:PSS based device. In addition, the value of series resistance for the CuO-based device has been calculated (by using the slope of the dark curve after 1.1 V) at Rs = 1.82 Ohm which is lower from the one calculated for PEDOT:PSS-based device at 3.12 Ohm. The above measurements show that the implementation of CuO-NPs HTLs significantly suppresses the parasitic resistances within the solar cell. These observations provide additional experimental evidences for the high hole-extraction ability of the proposed CuO-NPs HTLs for perovskite solar cells.

Fig. 4c shows the external quantum efficiency (EQE) as a function of wavelength for the two p-i-n devices under study. In accordance with Fig. 4a, the CuO NPs-based device exhibited higher EQE in almost all spectrum range (400-800 nm) compared to PEDOT:PSS-based device. The integrated Jsc values extracted from the EQE spectra of the two devices demonstrate a good match with the measured Jsc from the J-V characteristics, with a ~5 % and ~3 % deviation for the PEDOT:PSS-based and the CuO NPs-based devices, respectively. Despite the similar shape in the EQE spectra for both devices under study before 530 nm, the EQE spectra of the PEDOT:PSS-based device exhibits a significantly sharper drop compared with the EQE spectra of the CuO NPs-based device in the range of 530-630 nm. This observation could be partially explained by the higher transmittance of the CuO NPs electrodes in this region (Fig. 2a).

Fig. 4c shows the transmittance measurements of ITO/CuO or PEDOT:PSS/$CH_3NH_3PbI_3$ as well as of the full stacks under comparison ITO/CuO or PEDOT:PSS/$CH_3NH_3PbI_3$/PCBM/AZO without the Al back contact of the device (Fig. 4c). Importantly the transmittance of CuO-NPs/$CH_3NH_3PbI_3$ is higher compared with the transmittance PEDOT:PSS/$CH_3NH_3PbI_3$ after 600 nm. The CuO-NPs based full layer stack also show higher transmittance in the range after 600-800 nm compared with PEDOT:PSS-based layer stacks. It is likely the transmittance enhancement above 600 nm originates from CuO-NPs light scattering effects. We have recently shown that higher EQE is observed in the region of 550-760 nm when aluminum-doped zinc oxide nanoparticulate electron selective contacts are used within the n-part of the p-i-n perovskite PV structure, resulting in light scattering effects and increased back electrode reflectivity.[7] We propose that CuO nanoparticulate hole selective contacts further contribute to light scattering phenomena within the p-i-n perovskite PV structure. Thus, the higher EQE observed in the region of 550-800 nm for the perovskite solar cells incorporating CuO nanoparticulate hole selective contacts is attributed to additional scattered light from CuO NPs that result to increased reflective light of Al and, therefore, to improved absorption properties of the corresponding perovskite active layer.

**Conclusions**

We have demonstrated a novel solvothermal synthetic route for the development of room temperature solution processed nanoparticulate interfacial layers for highly efficient p-i-n perovskite solar cells. The CuO NPs and thin films are synthesized using thermal decomposition of CuCl diluted in DMSO at low temperatures and short reaction times. The synthetic procedure results in high purity monodispersed, spherical CuO NPs with monoclinic crystal structure and tunable sizes from 5 to 10 nm by simply changing the reaction temperature. The high quality CuO NP dispersions in DMSO were used to fabricate CuO hole selective contacts between ITO and Perovskite active layers with a thickness of about 15 nm without any post deposition temperature treatment. The use of CuO NPs as interfacial layer between ITO and $CH_3NH_3PbI_3$ active layer leads in increased Voc of 1.09 V and PCE of 15.3%, compared to that achieved with PEDOT:PSS-hole selective contact based $CH_3NH_3PbI_3$ solar cell device, which exhibits Voc of 0.9 and PCE up to 10.6%. The improved device performance arises from the bottom electrode CuO NPs-based interface modification. The quality of the proposed ITO/CuO-NPs bottom electrode is indicated by XPS/UPS measurements and the efficient quenching of the photoluminescence of $CH_3NH_3PbI_3$ films. In addition, the $CH_3NH_3PbI_3$ photoactive layer when fabricated on top of ITO/CuO NPs shows improved crystallinity and absorption characteristics which also contribute to the overall increase in device performance. Furthermore, improved light manipulation within the p-i-n $CH_3NH_3PbI_3$ solar cell is also observed due to nanoparticulate nature of the CuO interfacial layer. The above experimental observations suggest that nanoparticulate based interface modification provides a series of benefits in p-i-n perovskite solar cell operation. The solvothermal synthesis route proposed in this work could potentially be used for the preparation of other low temperature solution processed nanoparticulate metal oxide thin films suitable for flexible perovskite solar cell and other printing electronics applications.

## Conflicts of interest

There are no conflicts of interest to declare

## Acknowledgements

This project has received funding from the European Research Council (ERC) under the European Union's Horizon 2020 research and innovation programme (grant agreement No 647311).

Notes and References

# Supplementary Information

**Room Temperature Nanoparticulate Interfacial Layers for Perovskite Solar Cells via solvothermal synthesis.**

Achilleas Savva[a], Ioannis T. Papadas[a], Dimitris Tsikritzis[a], Gerasimos S. Armatas[b], Stella Kennou[c] and Stelios A. Choulis[a,†]

[a]Molecular Electronics and Photonics Research Unit, Department of Mechanical Engineering and Materials Science and Engineering, Cyprus University of Technology, Limassol, 3603, Cyprus.

[b]Department of Materials Science and Technology, University of Crete, Heraklion 71003, Greece.

[c]Department of Chemical Engineering, University of Patras, Patra, 26504 Greece

† stelios.choulis@cut.ac.cy

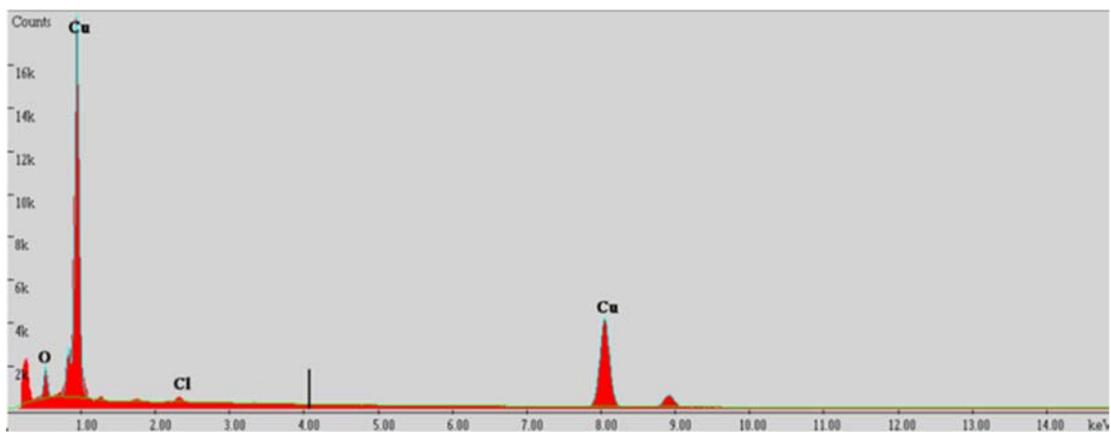

**Fig. S1:** EDS analysis of the as prepared CuO NPs obtained at 100 ºC.

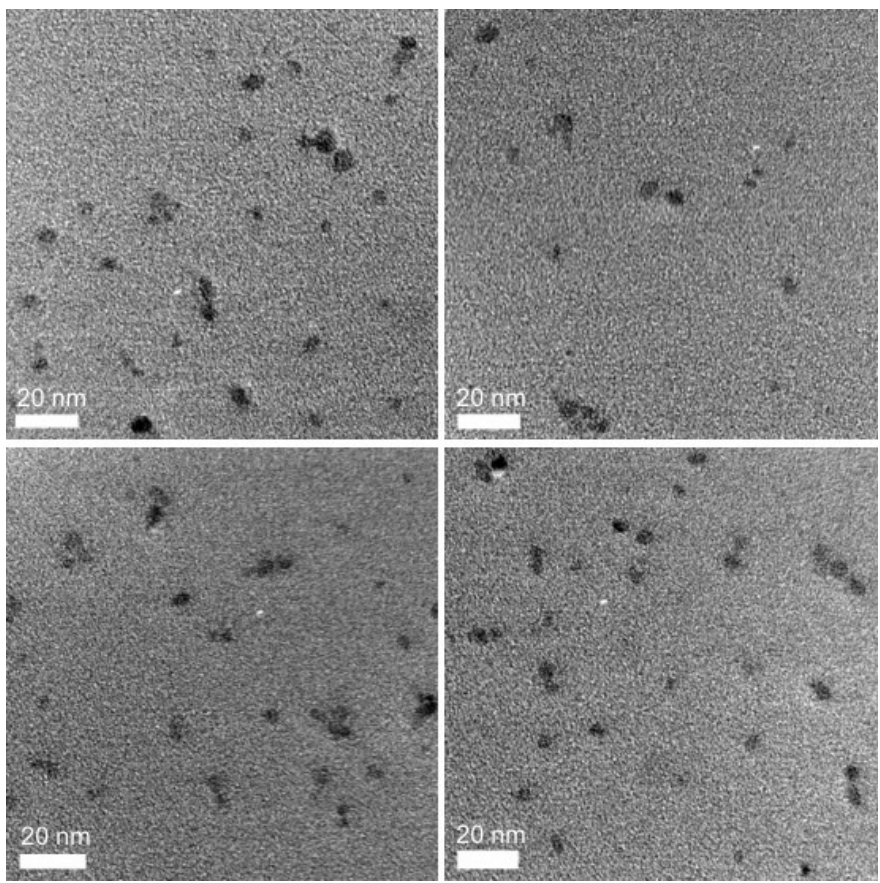

**Fig S2:** Representative TEM images for CuO NPs obtained at 100 ºC decomposition temperature. The average diameter of the CuO NPs was estimated by counting more than 100 individual NPs in several TEM images. It should be stressed that the processing of the sample for TEM imaging was based on the requirements of the TEM instrumental conditions for the observation of nanoparticles (see Experimental section within the article) and is not directly related to the processing conditions used on the preparation of CuO interfacial layers for perovskite solar cells.

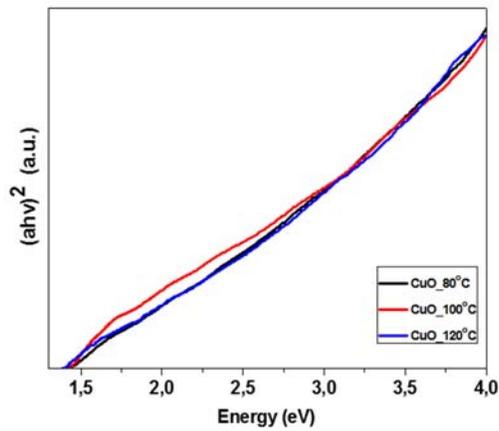

**Fig. S3:** The $(\alpha h v)^2$ vs. $hv$ plot for CuO nanoparticulate powders synthesized at 80 ºC (black), at 100 ºC (red) and at 120 ºC (blue line).

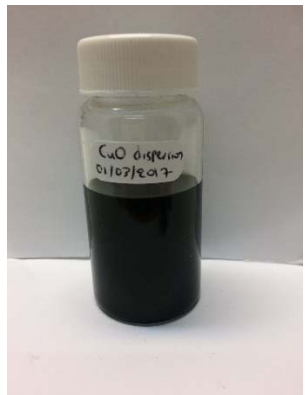

**Fig. S4:** Digital photograph of CuO dispersions in DMSO, 20 mg/ml. The dispersions where stable several months after fabrication. The concentrated dispersions where then diluted down to 0.5 mg/ml for the fabrication of CuO HTLs in the range of ~15 nm.

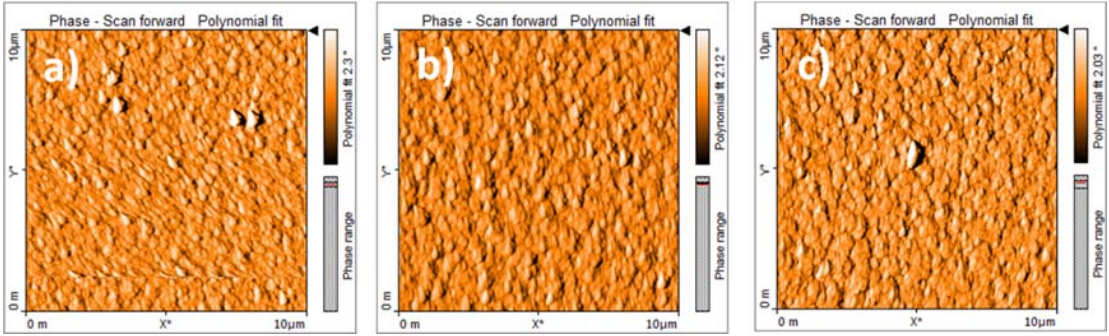

**Fig. S5:** AFM measurements for CuO-NPs thin films spin coated from DMSO despersions at 0.5 mg/ml without any post deposition treatment. **a)** CuO NPs synthesized at 80 °C, **b)** CuO NPs synthesized at 100 °C **c)** CuO NPs synthesized at 120 °C.

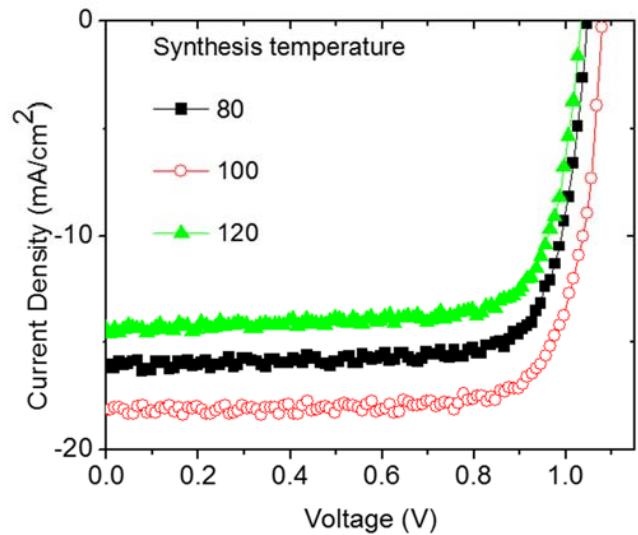

**Figure S6:** Current density–Voltage (J-V) characteristics, for p-i-n solar cells comprised of ITO/CuO-NPs/CH$_3$NH$_3$PbI$_3$/PC[70]BM/AZO/Al. CuO NPs synthesized at 80 °C with average size 5 nm (black filed squares), CuO NPs synthesized at 100 °C with average size

6 nm (red open circles) and CuO NPs synthesized at 120 ºC with average size 9 nm (green filed triangles).

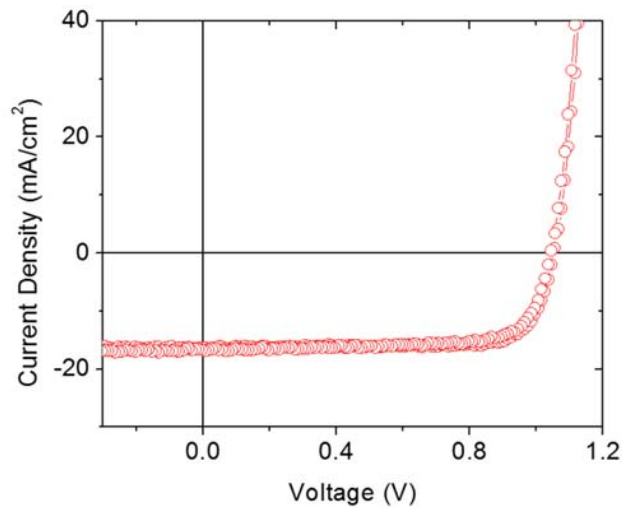

**Fig. S7:** Current density–Voltage (J-V) characteristics, for p-i-n solar cells comprised of ITO/CuO-NPs-100ºC/CH$_3$NH$_3$PbI$_3$/PC[70]BM/AZO/Al in forward (short circuit -> open circuit) and reverse (open circuit -> short circuit).

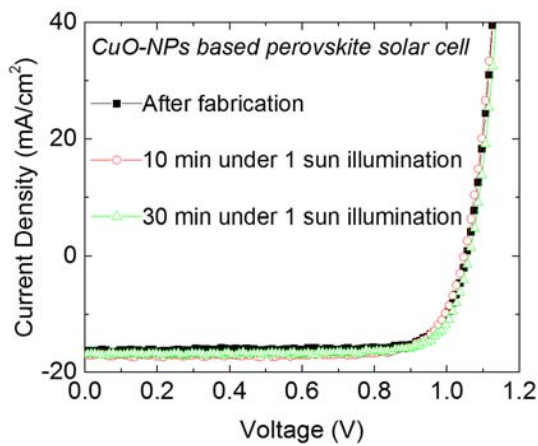

**Fig. S8:** Current density–Voltage (J-V) characteristics, for a representative CuO-based p-i-n solar cell, measured after fabrication (black filled squares), after 10 minutes under continues 1 sun ilumination (red open circles) and after 30 minutes of continuous 1 sun illumination (green open triangles).

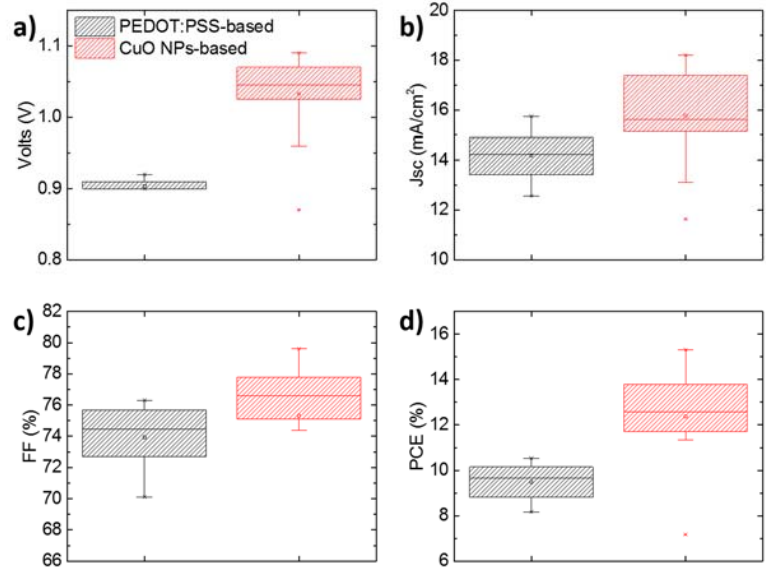

**Fig. S9:** Average photovoltaic parameters represented in box plots out of 16 devices of each series of p-i-n perovskite soalr cells under study. PEDOT:PSS-based represented with black box plots and CuO-based with red box plots. **a)** Open circuit voltage (Voc) **b)** current density (Jsc) **c)** fill factor (FF) and **d)** power conversion efficiency (PCE).